\documentclass[aps,pre,print,groupedaddress]{revtex4-1}
\usepackage{graphicx}  
\usepackage{dcolumn}  
\usepackage{bm}           
\usepackage{amsmath}
\usepackage{amsfonts}
\usepackage{amssymb}
\usepackage{color}
\usepackage[normalem]{ulem}
\usepackage{ wasysym }
\usepackage{dsfont}
\usepackage{mathtools}
\usepackage{amsthm}

\usepackage{array}
\newcolumntype{L}[1]{>{\raggedright\let\newline\\\arraybackslash\hspace{0pt}}m{#1}}
\newcolumntype{C}[1]{>{\centering\let\newline\\\arraybackslash\hspace{0pt}}m{#1}}
\newcolumntype{R}[1]{>{\raggedleft\let\newline\\\arraybackslash\hspace{0pt}}m{#1}}

\newtagform{AArabic}[]()

\usepackage{todonotes}

\begin{document}

\title{Turing patterns on a two-component isotropic growing system.\\ Part 1: Homogeneous state and stability of perturbations in absence of diffusion}

\author{Aldo Ledesma-Dur\'an}
\email{aldo\_ledesma@xanum.uam.mx}
\affiliation{UAM-Iztapalapa}

\date{11.~July 2023}

\begin{abstract}
The reaction-diffusion processes in a growing domain involves  a dilution term that modifies the properties of the homogeneous state that, in contrast to a fixed domain, depends on time. We study how the dilution term changes the steady concentrations and modifies the stability properties of the perturbations. We propose a solution for the homogeneous state that incorporates these factors and is valid for slow variation of the size of the domain which is based on a linear approach and has been tested against numerical solutions for different types of growing: exponential, linear, quadratic and oscillatory. We prove that the deviation of the steady state is proportional to the fixed point concentration, and occurs most notably for exponential growth. Systems with linear or quadratic growth tend to recover the state that would have in absence of diffusion, whereas the oscillatory variation of the domain size produce temporal oscillations  of the concentration. Regarding the Turing conditions for the apparition of spatial patterns, we study those related to stability in absence of diffusion and establish that in a growing domain it depends upon: the change in the steady state concentrations, the local change of volume that affects the concentration and the stability that the reactive system would have in absence of dilution. These conditions provide richer conditions for the emergence of patterns than those found in a fixed domain. The formal results provided in this work are verified against numerical simulations of the homogeneous state for the Brusselator and BVAM reactions and we discuss how these variations of the homogeneous state can give raise to crucial differences in the formation of Turing patterns in growing domains.
\end{abstract}

\maketitle

\section{General presentation of this series}

Turing patterns in reaction-diffusion systems have a wide range of applications related to the self-organization of biological and chemical systems where the domain changes  size \cite{knobloch2015problems,marcon2012turing}. Empirical evidence from numerical simulations shows that the shape of the pattern in an growing domain at a specific time depends on its past history \cite{okuda2018combining,konow2019turing}. This phenomenon, which can be understood as a type of memory, is probably related to persistence, \emph{i.e.}, the ability of a dissipative structure to maintain a wavenumber, even when there is another wavenumber that may be more stable (more resistant to sideband perturbances) \cite{krechetnikov2017stability,ledesma2020eckhaus,ghadiri2019pattern}. 

To test this possibility, a non-linear approximation to the solution of the reaction-diffusion-dilution (RDD) system near the Turing bifurcation is required. This, in turn, requires establishing the conditions of the linear system for the appearance of Turing patterns in growing domains which remains an open problem. Finding the Turing bifurcation has been an elusive problem due to the temporal dependence of the coefficients involved in the RDD system, and many attempts and progress have been made without being completely conclusive \cite{ madzvamuse2010stability,van2021turing}.

The strategy followed in this series of papers differs from previous attempts where approximations for the linear problem are sought and then conditions for its stability are determined. \cite{kim2020pattern,vittadello2021turing}. Since we will focus on a two-component system, we will try to find stability conditions based on the phase plane structure. This will lead us to the problem of finding a type of potential function in which all trajectories decrease to a stable point in the absence of diffusion, and some are saddle fixed points for some wavenumber when diffusion is on. We will show that the changes in the structure of the potential function allows us to establish some hypotheses about when the Turing pattern emerges. These hypotheses will be tested against numerical simulations of reaction diffusion systems using the finite elemen method in a one-dimensional RDD system with different types of growth/shrinkage. In particular, we will focus on linear, quadratic, exponential, and oscillatory growth functions.

Throughout this work, we will highlight some aspects related to the physical-chemical interpretation of the processes. For example, we will distinguish two magnitudes that intervene in the description of the problem: the chemical concentration of the agents and their molar quantity. Unlike a fixed domain, in a domain that varies with time, the stability/instability of one quantity does not imply the stability/instability of the other due to the dilution effect. Another fact that leads to different conditions for the Turing bifurcation is the steady state concentration of the reaction since, if the reactor is in chemical equilibrium with a reservoir of non-zero concentrations, the effect of dilution is also to change the internal chemical flux resulting in a new steady concentration, in contrast to a reactor with an isolated reactor without constant external flow.

In this series of works we will study the homogeneous state that, unlike a fixed domain, usually depends on time. We will study the Turing conditions for exponential growth and develop both the linear and non-linear approximation of the perturbations, showing the main features of the Turing conditions in growing domains. This will allow us to study aspects related to persistence  and hysteresis as a non linear responses sideband disturbances. Given this example as a touchstone, we will outline our strategy for finding the Turing conditions by studying the  phase space's structure. We will compare our hypothesis with the exponential case to test its possible validity and illustrate the different particular aspects of other types of growth, highlighting the differences with fixed domains. Finally we will present the possibility of understanding time as a type of bifurcation parameter that indicates when a pattern can appear or disappear from a system solely due to the dilution effect.

\subsection{The problem of the homogeneous state}
The conditions for the appearance of Turing patterns are related to the response of a reaction-diffusion system to a small spatial perturbation. However, in a domain that increases with time, the dilution term can produce a homogeneous state where chemical concentrations can vary with time. Unlike a domain of fixed size, in an increasing domain, the response to disturbances combines the local production/consumption determined by reaction parameters with the local change in volume introduced by the dilution term.

For this reason, before considering the conditions for the occurrence of Turing patterns in an increasing domain, we will first discuss the properties of the homogeneous state of an RDD system, \emph{i.e.}, the state not related to the spatial dependence and on which the spatial perturbations are expected to increase or decrease depending on the parameters of the RDD system. As we shall see, the dilution term itself introduces direct changes in fixed-point concentrations for exponential growth, temporal variations for polynomial and oscillatory growth, and even, for some parameter values, temporal oscillations for shrinking domains.

We will discuss in this article how these changes not only affect the steady state, but can also change the dynamic properties of the disturbances themselves and thus can give rise to new situations related to the appearance/disappearance of Turing patterns that do not occur in a fixed domain. For our work we will consider the simple situation of a two-component RDD system with constant coefficients under isotropic growth. To exemplify our predictions, we will use the well-known chemical systems of the Brusselator \cite{pena2001stability} and the BVAM model \cite{barrio2008turing}.

For this Part 1 of the work, the structure is as follows. In Section \ref{sec:presentation}, we present a derivation of the RDD equations based on simple physical terms for one-dimensional domains with isotropic growth. In Section \ref{sec:homogeneous}, we provide formal solutions for the homogeneous state based on a linear approximation. It also discusses when this time-dependent solution can be approximated by a constant solution for the exponential, linear quadratic, and oscillatory cases. In Section \ref{sec:brusselas}, we exemplify this hypothesis for the Brusselator reaction system and provide numerical criteria for the validity of our approximation. Based on these considerations, in Section \ref{sec:perturbations}, we discuss how the combined effect of steady-state change and dilution can inherit or change the stability properties of perturbations respect to that of a fixed domain, and in Section
\ref{sec:bvam} we compare with a system with a fixed point at the origin. Finally, in Section \ref{sec:discussion}, we present a brief summary of conclusions and discuss the perspective of our results.

\section{Equations for the  homogeneous state and perturbations of a RDD system}\label{sec:presentation}

First of all, we will study the differences between the homogeneous state and those modes dependent on the spatial coordinate commonly understood as perturbations. To do so, we will first derive the RDD equation that describes the interaction of reaction, diffusion, and dilution processes in growing domain. 
 
Let us consider a one-dimensional domain that grows homogeneously with time $t$ and has size $l(t)$ on a straight line. At time $t$, we will quantify the mass between the position $x_l(t)$ on the left and the far right which is at $x_r(t)$. For a species, if its density is $c(x,t)$ (in units of number of particles per unit of length), the diffusive flux (in units of number of particles per unit of time) is $j=-D \partial c / \partial x$, where $D$ is its diffusion coefficient. Conservation of mass implies that the number of molecules in the time interval $t$, \emph{i.e.} $\hat{n}(t)= \int \limits_{x_l}^{x_r} c( x,t)\ ,dx$, would change due to the entry or exit of material through the boundaries and internal production/consumption:

\begin{equation}\label{eq:conservation1}
\frac{d}{dt} \int \limits_{x_l}^{x_r} c(x,t)\,dx =-[ j(x_r)- j(x_l)]+ \delta F(t),
\end{equation}
\noindent where $\delta F(t)$ is the production/consumption of molecules between $x_l$ and $x_r$ (in units of particles produced/consumed per unit time). Using Leibniz's integral rule on the left and  substitution of the flux on the right lead to

\begin{equation}
\int \limits_{x_l(t)}^{x_r(t)} \frac{\partial c}{ \partial t}  (x,t)\,dx+ c (x_r, t ) \dot{x}_r(t) - c (x_l, t ) \dot{x}_l(t)=D \left(  \frac{ \partial c}{ \partial x} \biggr \rvert _{x_l} -  \frac{ \partial c}{ \partial x} \biggr \rvert _{x_0} \right) + \delta F.
\end{equation}
\noindent The first term at the lhs represents the local change in concentration in the interval from $x_l(t)$ to $x_r(t)$, and the other two represent the difference in the number of particles with respect to the new size ranging from $x_l (t+\delta t)$ to $x_r(t+ \delta t)$. The terms at the lhs are the number of particles produced locally in the spatial interval and in an infinitesimal time $\delta t$. Dividing by $\delta x=x_r -x_l$ and taking the limit as this difference approaches $\delta x\to 0$, the above equation can be approximated by

\begin{equation}\label{eq:rd1d}
 \frac{\partial c}{ \partial t}  + \frac{\dot{l}(t)}{l(t)} c (x, t )  = D   \frac{ \partial^2 c}{ \partial x^2}  + f(t).
\end{equation}
In this equation, we have used that $\delta x^{-1}\,d(\delta x)/\delta t \to \dot{l}(t)/l(t)$ is the rate of change of growth representing how many size units (in ) the domain gains/loses per unit of time. This dilution term quantifies the local change in concentration due to growth. In addition, we have used that the reaction locally satisfies $\delta F/\delta x \to f$, and therefore, $f(t)$ is the production/consumption of particles per unit length and time and, therefore, is a local measure of the reaction that does not depend on the size of the domain.

If the volume change is homogeneous (isotropic growth), the relationship between the moving and fixed reference frame with coordinates $\xi$ is $ x =x_0 +l(t) \xi$ with $\xi \in [0, 1]$. In this case, the last equation in terms of the fixed coordinate is

\begin{equation}\label{eq:rd1dfijo}
 \frac{\partial c}{ \partial t}  +   \frac{\dot{l}(t)}{l(t)} c (\xi, t ) = \frac{D}{l^2(t)}   \frac{ \partial^2 c}{ \partial \xi^2}  +f(c,t).
\end{equation} 
Eq. \eqref{eq:rd1dfijo} is just one case of RDD equations applicable to one-dimensional, Euclidean, and isotropically growing domains. Other deductions can be made by different methodologies and in more general frameworks including curved domains or anisotropic growth \cite{crampin1999reaction,plaza2004effect,krause2019influence}.  

The equation RDD \eqref{eq:rd1dfijo} can be generalized to various components. If $\mathbf{c}$ represents the concentration vector with diagonal diffusion matrix $ \mathds{D}$, and the reaction vector is given by $\mathbf{f}$, the equation that describes the reaction-diffusion dynamics in a domain with isotropic growth is

\begin{equation}\label{eq:system}
 \frac{\partial  \mathbf{c}}{ \partial t}  +   \frac{\dot{l}(t)}{l(t)} \mathbf{c}(\xi,t) = \frac{1}{l^2(t)}   \mathds{D}  \frac{ \partial^2  \mathbf{c}}{ \partial \xi^2}  +\mathbf{f}( \mathbf{c}).
 \end{equation} 
where the relation between the actual and fixed coordinates is  $ x =x_0 +l(t) \xi$  with $\xi \in [0,1]$.

To exemplify the different results of this series of works, we use the Brusselator and BVAM reactive systems, whose linear/nonlinear approach near the Turing bifurcation in fixed domains is well known. \cite{pena2001stability,meixner1997generic,ledesma2019primary}. If $\mathbf{c}=(c_u,c_v)^T$, they are given respectively by :

\begin{equation}\label{eq:bruselas1}
 \mathbf{f}(\mathbf{c})=(A-Bc_u-c_u +c_u^2 c_v, Bc_u -c_u^2 c_v)^T,
\end{equation}

\begin{equation}\label{eq:bvam1}
 \mathbf{f}(\mathbf{c})=(c_u+a c_v-  c_u c_v-c_u c_v^2 , b c_v+ h c_u+ c_u c_v+ c_u c_v^2)^T.
\end{equation}
These equations follow from the law of mass action. The Brusselator is an example of open reactors where a reactant that produces the activator is held constant, leading to a non-equilibrium steady state where the catalyst concentrations are non-zero. The BVAM model, under appropriate conditions, has a single fixed point at the origin and can be understood to reflect a closed reactor where concentrations will tend to zero. As we will see, this difference would affect the ability of an RDD system to change its fixed-point concentration in the long term and, therefore, the stability of perturbations.

To understand this, we note that when studying the Turing bifurcation in fixed domains, the stability of spatial perturbations is considered around a fixed-point concentration $\mathbf{c}_0$, defined by setting all reactions equal to zero, $ \mathbf{ f}(\mathbf{c}_0)=\mathbf{0}$. In open systems, this fixed point is usually not the origin, but a constant state defined by $\mathbf{c}_0$ to which the concentrations will tend asymptotically. For a growing domain, if we consider a perturbation $\boldsymbol{\zeta}(\xi,t) =\mathbf{c}(\xi,t)-\mathbf{c}_s(t)$ around a homogeneous state $\mathbf{c }_s(t)$, from eq. \eqref{eq:system} it follows that

\begin{equation}\label{eq:fullseparation}
 \frac{\partial   \mathbf{c}_s}{ \partial t}
  +   \frac{\dot{l}(t)}{l(t)} \mathbf{c}_s-
   \mathbf{f}( \mathbf{c}_s)=
  - \frac{\partial \boldsymbol{\zeta}}{ \partial t} - \frac{\dot{l}(t)}{l(t)} \boldsymbol{\zeta}  +
   \frac{1}{l^2(t)}   \mathds{D}  \frac{ \partial^2  \boldsymbol{\zeta}}{ \partial \xi^2} + 
  \frac{\partial \mathbf{f}}{\partial \mathbf{c}}(\mathbf{c}_s)\boldsymbol{\zeta} + \mathcal{O}(\boldsymbol{\zeta}^2).
\end{equation} 
Neglecting quadratic terms in the perturbations, this leads to two problems. The homogeneous state obeys

\begin{equation}\label{eq:homogeneous}
 \frac{\partial   \mathbf{c}_s}{ \partial t} +\frac{l'(t)}{l(t)} \mathbf{c}_s=\mathbf{f}(\mathbf{c}_s), 
\end{equation}
whereas, using this solution to evaluate $\frac{\partial \mathbf{f}}{\partial \mathbf{c}}(\mathbf{c}_s)$, the perturbation $\boldsymbol{\zeta}$ obeys

 \begin{equation}\label{eq:perturbation}
 \frac{\partial \boldsymbol{\zeta}}{ \partial t}   +\frac{\dot{l}(t)}{l(t)}\boldsymbol{\zeta} = \frac{1}{l^2(t)}   \mathds{D}  \frac{ \partial^2  \boldsymbol{\zeta}}{ \partial \xi^2}  + \frac{\partial \mathbf{f}}{\partial \mathbf{c}}(\mathbf{c}_s) \boldsymbol{\zeta}.
\end{equation} 

Equation \eqref{eq:homogeneous} is a nonlinear problem that in general cannot be solved analytically. The solution will depend on time, which would affect the evaluation of $\frac{\partial \mathbf{f}}{\partial \mathbf{c}}(\mathbf{c}_s)$, in turn modifying the stability of the perturbations in \eqref{eq:perturbation}. Therefore, before tackling this problem, let us consider in the next section a linear approximation for the solution of the homogeneous state \eqref{eq:homogeneous} to understand its properties.

\section{Approximation to the homogeneous state}\label{sec:homogeneous}

To approximate the solutions of a non-linear problem such as \eqref{eq:homogeneous}, the type of expected solution must be taken into account. Since we are looking for Turing patterns, disturbances need to be stable in the absence of diffusion. Therefore, it is expected that this stability is inherited from the stability of the homogeneous state. We will see later that this is not always the case, that is, there can be stable homogeneous states that give rise to unstable disturbances and vice versa. However, in most cases, we have observed that if the first is asymptotically stable, the same is true for the other and, therefore, its behavior can be predicted using a linear approach.

\subsection{Time dependent and stable homogeneous state }\label{sec:homotime}
  
Let's find an approximate solution of \eqref{eq:homogeneous} assuming that the concentration values are already close to the fixed-point concentration $\mathbf{c}_0$ where $\mathbf{f}(\mathbf{c} _0) =\mathbf{0}$. In this case, the reaction can be approximated as $\mathbf{f}(\mathbf{c}_s)\approx \mathds{J} [\mathbf{c}_s(t)-\mathbf{c}_0]$ where $\mathds{J}$ is the constant Jacobian matrix of the reaction evaluated at $\mathbf{c}_0$. Defining the deviations around the homogeneous state as $\delta \mathbf{c}(t)=\mathbf{c}_s(t)-\mathbf{c}_0$, this leads to
 
\begin{equation}\label{eq:st1steadystatelinear}
 \frac{d\,[\delta \mathbf{c}]}{dt}+\frac{l'(t)}{l(t)}[\mathbf{c}_0 +\delta \mathbf{c}]\approx \mathds{J}\, \delta \mathbf{c}.
\end{equation}

If $\mathds{J}$ is a diagonalizable constant square matrix with modal matrix $\mathds{P}$ such that $\boldsymbol{\Lambda}=\mathds{P}^{-1}\mathds{J}\mathds{P}$, then the solution of \eqref{eq:st1steadystatelinear} with initial condition $\delta \mathbf{c}(0)$ is

\begin{equation}\label{eq:st1perturbation1}
\delta \mathbf{c}=  \frac{l(0)}{l(t)} \mathds{P}  e^{\boldsymbol{\Lambda} t} \mathds{P}^{-1} \delta \mathbf{c}(0)-
 \mathds{P} \frac{e^{\boldsymbol{\Lambda} t}}{l(t)} \left(\int \limits_0^t  l'(t') e^{-\boldsymbol{\Lambda} t'}dt' \right)\mathds{P}^{-1}  \mathbf{c}_0.
\end{equation}
with $\boldsymbol{\Lambda}$ the diagonal matrix whose entries are the eigenvalues of $\mathds{J}$.  

To show this, let us multiply  \eqref{eq:st1steadystatelinear} by $l(t)$ and define $ \mathbf{z}=l(r) \mathds{P}^{-1} \delta \mathbf{c}$. This  leads to $\mathds{P}\dot{\mathbf{z}} -\mathds{J} \mathds{P} \mathbf{z}=- l'(t) \mathbf{c}_0 $ since $\mathds{P}$ is also a constant matrix (because $\mathds{J}$ is constant). Multiplying both sides for $\mathds{P}$ and using the diagonal matrix $\boldsymbol{\Lambda}$, the resulting equation for the i-esim component is decoupled from the others: $\dot{z}_i- \Lambda_{i} z_i=-l'(t) [\mathds{P}^{-1} \mathbf{c}_0 ]_i$, where $\Lambda_i \equiv \boldsymbol{\Lambda}_{ii}$ are the eigenvalues of $\mathds{J}$, which  we here assumed to be all different to simplify the proof.  This first-order equation is solved using the integrating factor method and leads to

\begin{equation}
z_i(t)= e^{\Lambda_{i} t} z_i(0)  -   e^{\Lambda_i t}  \left(\int \limits_0^t  l'(t') e^{-\Lambda_i t'}dt' \right)  [\mathds{P}^{-1} \mathbf{c}_0 ]_i.
\end{equation}
Noticing then than $ e^{\Lambda_i t}=\sum_j [e^{\boldsymbol{\Lambda} t}]_{ij}\delta_{ij} $, $z_i(0) =l(0)\sum_k \mathds{P}_{ik}^{-1} \delta \mathbf{c}_k(0)  $ and  $  [\mathds{P}^{-1} \mathbf{c}_0 ]_i=\sum_m  \mathds{P}^{-1}_{im} [\mathbf{c}_0]_m$, the above equations can be written in a single matrix expression as 

\begin{equation}
\mathbf{z}=l(0) e^{\boldsymbol{\Lambda t}} \mathds{P}^{-1} \delta \mathbf{c}(0)   -   e^{\boldsymbol{\Lambda}t}  \left(\int \limits_0^t  l'(t') e^{-\boldsymbol{\Lambda} t'}dt' \right)  \mathds{P}^{-1} \mathbf{c}_0,
\end{equation}
which becomes \eqref{eq:st1perturbation1} using $\delta \mathbf{c}=l^{-1}(t) \mathds{P} \mathbf{z}$. This completes the proof $\square$ .

\vspace{0.35cm}

The Eq. \eqref{eq:st1perturbation1} states that the deviation of the homogeneous state $\mathbf{c}_s$ from the fixed-point concentration  $\mathbf{c}_0$ has two terms:

\begin{itemize}
\item A contribution proportional to the initial deviation $\delta \mathbf{c}(0) $ that will decay exponentially to zero if all eigenvalues of $\mathds{J}$ satisfy $\mathcal{R}e[\Lambda_i] t- \ln l(t) / l(0)<0$. This term represents the competition between the reaction and the dilution effect through the ratio $e^{\Lambda_i t}/l(t)$ that can rise or fall depending on the type of growth.

\item A contribution proportional to the fixed-point concentration $\mathbf{c}_0$. The magnitude of this deviation is proportional to the factor $e^{\Lambda_i t} \left(\int \limits_0^t l'(t') e^{-\Lambda_i t'} dt' \right)/l(t)$ and, as we will show with our examples, is more pronounced at the beginning of the process for growth and towards the end for shrinkage.
\end{itemize}

From the result in \eqref{eq:st1perturbation1}, assuming that the system is already initially at the fixed-point concentration $\mathbf{c}_0$ and that the nonlinear terms of \eqref{eq:homogeneous} are negligible, the homoheneous state obeys approximately.

\begin{equation}\label{eq:timeaproxima}
\mathbf{c}_s \approx  \left[\mathds{I}-   \mathds{P} \frac{e^{\boldsymbol{\Lambda} t}}{l(t)} \left(\int \limits_0^t  l'(t') e^{-\boldsymbol{\Lambda} t'}dt' \right)\mathds{P}^{-1}   \right] \mathbf{c}_0.
\end{equation}
Later in our numerical examples we will see what factors contribute to the nonlinearity of the problem, and therefore to the failure of \eqref{eq:timeaproxima} as an approximation.

\subsection{Explicit computation for some types of growing}

To make some progress, the expression \eqref{eq:timeaproxima} can be computed explicitly for some functions $l(t)$ along these general lines: Given $l(t)$ as an exponential, sinusoidal, or polynomial function, the integrals of the forms $e^{\Lambda_i}/l(t) \int_0^t e^{\Lambda_i t} l'(t') dt' $ can be computed explicitly yielding rational functions similar to $ l (t)$ in numerator and denominator, and in terms of powers of $\Lambda_i^m$. Since all matrices involving $\boldsymbol{\Lambda}$ in \eqref{eq:timeaproxima} are diagonal, the scalar result can be converted to a diagonal matrix using the direct replacement of $\Lambda_i \to \boldsymbol{\Lambda}$. Then using the properties of the modal matrix, all terms of the matrix $\mathds{P}\boldsymbol{\Lambda}^{m}\mathds{P}^{-1}$ are replaced by terms of the form $ \mathds{J}^{m}$. This allows us to find the explicit expression for $\mathbf{c}_s$. Let's better illustrate this cumbersome explanation with an example.

\vspace{0.5cm}

\paragraph*{Example: Quadratic growth.-} Let $l(t)=l_0(1+r t^2)$. For this case, we can explicitly calculate for any $i$:

\begin{equation}
\frac{e^{\Lambda_i t}}{l(t)}\int\limits_0^t  e^{-\Lambda_i t'} l'(t')dt' =\frac{2 r \left(-t \Lambda_i  +e^{\Lambda_i  t}-1\right)}{\Lambda_i ^2 \left(1+r t^2\right)}.
\end{equation}
assuming that all eigenvalues are non zero. Since all involved matrices in \eqref{eq:timeaproxima} are diagonal, this means that

\begin{equation}
\frac{e^{\boldsymbol{\Lambda} t}}{l(t)}\int\limits_0^t  e^{-\boldsymbol{\Lambda} t'} l'(t')dt' =\frac{2 r}{1+r t^2}  \left(-t \boldsymbol{\Lambda}^{-1}  +\boldsymbol{\mathds{J}}^{-2} e^{\boldsymbol{\Lambda} t}-\boldsymbol{\Lambda}^{-2}\right).
\end{equation}
 Using systematically the property $\mathds{J}^m=\mathds{P}\boldsymbol{\Lambda}^m \mathds{P}^{-1}$ (which implies $e^{\mathds{J}}=\mathds{P}e^{\boldsymbol{\Lambda}} \mathds{P}^{-1}$ ), from \eqref{eq:timeaproxima} we obtain that:

\begin{equation}
\mathbf{c}_s =  \left[\mathds{I}- \frac{2 r}{1+r t^2}  \left(-t \mathds{J}^{-1}  +\boldsymbol{\mathds{J}}^{-2}e^{\mathds{J} t}-\boldsymbol{\mathds{J}}^{-2}\right) \right] \mathbf{c}_0.
\end{equation}

\vspace{0.5cm}

\begin{table*}[h!]
\begin{tabular}{|C{4cm}|C{11cm}|}
\hline 
Size ratio $l(t)/l(0)$ & Deviation $ \boldsymbol{\delta C}(t)$ \\ 
\hline 
\hline
$e^{r t}$  & $\displaystyle r (\mathds{J} -r \mathds{I})^{-1} (\mathds{I} -e^{\mathds{J}t} e^{-rt} )$  \\ 
\hline 
 $1+r t $ & $\displaystyle \frac{r}{1+ r t}  \mathds{J}^{-1} \,\left( \mathds{I}-e^{\mathds{J}t}  \right)$ \\ 
 \hline
  $1+r t^2 $ & $\displaystyle \frac{2r}{1+ r t^2}  \mathds{J}^{-2} \,\left( \mathds{I}-e^{\mathds{J}t} +t \mathds{J}  \right)$ \\ 
\hline
 $1+r \sin (\omega t ) $ & $\displaystyle  \frac{r \omega}{1+ r \sin(\omega t)}  (\mathds{J}^2+ \omega^2 \mathds{I})^{-1} [ \mathds{J} \cos(\omega t )- \omega \sin(\omega t)-\mathds{J} e^{\mathds{J}t} ]   $ \\ 
\hline
\end{tabular} 
 \caption{The temporal deviation of the homogeneous state from  the fixed point $\mathbf{c}_0$ in \eqref{eq:deviation2}  for different types of growth in the first column.  }\label{tab:deviations}
 \end{table*}
 
From \eqref{eq:timeaproxima}, in general we have that

 \begin{equation}\label{eq:deviation2}
 \mathbf{c}_s -\mathbf{c}_0 =  \boldsymbol{\delta C}(t)\, \mathbf{c}_0,
\end{equation} 
and the form of this matrix $\boldsymbol{\delta C}(t)$ that measures the deviation between homogeneous state and fixed-point concentration is given in Table \ref{tab:deviations} for the types of growth in the first column. For the cases summarized in Table \ref{tab:deviations}, the explicit functional form of the deviation in \eqref{eq:deviation2} allows us to glimpse some properties of the homogeneous state:

\begin{itemize}

 \item \emph{General traits.-} The deviation between the homogeneous state and the fixed point is proportional to $r$, the parameter measuring for each case the variation of the size of the domain. It can also be noted that the damping of the solution depends on the eigenvalues of the Jacobian matrix evaluated at $\mathbf{c}_0$ through the exponential factor $e^\mathds{J}$, so the stability of the homogeneous state partially inherits the local stability of the chemical reaction.Besides, the deviation is proportional to some term that depends on the negative powers of $\mathds{J}$ (or some modification), which means that if the eigenvalues of $\mathds{J}$ are closer to zero (near bifurcation), the $\boldsymbol{\delta C}(t)$ deviations are larger.

 \item \emph{Particular traits.-}  For the exponential case, the linear deviation of the fixed point of concentrations for long times is proportional to the constant values $r/(r-\Lambda_i)$, so the steady state can be different from the fixed point. For linear growth, if the eigenvalues of $\mathds{J}$ have a negative real part, the deviation decays to zero as $r / (1+ r t )$, and something similar occurs for quadratic growth with rate $ rt /( 1+ rt^2)$; therefore in these two cases, in an increasing domain ($r>0$) the deviations tend to $0$, while for the decreasing situation ($r<0$), the deviation grows over time. For oscillatory growth, the deviations maintain the oscillatory behavior with an amplitude proportional to $r\omega$; furthermore, the amplitude of the deviations is affected by the distance between the frequency of the domain oscillation (measured by $\omega$) and the frequency of the resulting chemical reaction (measured by the imaginary part of the eigenvalues of $\mathds{J}$).  

\end{itemize}

All these features will be corroborated by our numerical simulations in the next section. However, the most important thing to note now is that many of the scenarios presented here derive in situations where the homogeneous state tends to a constant value (exponential growth/shrinkage) or tends to fixed-point concentration (linear or quadratic growth) or maintains low amplitude oscillations around a concentration (oscillatory case). In the next section we will present the idea of using a representative value of the homogeneous concentration in order to simplify the time dependence of the Jacobian of the problem in \eqref{eq:perturbation}.

\subsection{ Representative time independent homogeneous state}

For the cases summarized in the last paragraph, after a transient time, it can happen that the solutions of homogeneous state remain basically constant during an interval of time. It can then be thought that an appropriate and constant value of the concentrations, call it $\mathbf{C}^0$, can capture the main qualitative characteristics of the dynamics. In a two-component system, for example, the constant solution would be expected not to significantly change the trace and the determinant of the Jacobian matrix. This would simplify the \eqref{eq:perturbation} problem, since now the linear reaction term $\frac{\partial \mathbf{f}}{\partial \mathbf{c}} (\mathbf{C}_0)$ does not depend on time. In this section we will pose this possibility.

Assuming that we are studying the Turing formation process in an interval $T=t_f-t_i$. If $\mathbf{c}(t)$ does not change much in $T$, a characteristic value of the concentrations can be their time average. From \eqref{eq:deviation2}  
  
 \begin{equation}\label{eq:representativo}
 \mathbf{C}_0  = \mathbf{c}_0 +\frac{1}{t_f-t_i}  \left( \int \limits_{t_i}^{t_f }\boldsymbol{\delta C}(t) \, dt \right) \mathbf{c}_0,
\end{equation} 
The formal computation of these time integrals for most $l(t)$ functions involves special functions whose complexity adds nothing to the physical idea and therefore, in our work they will be performed numerically.

The exception is integral growth where \eqref{eq:representativo} leads to
\begin{equation}
 \mathbf{C}_0  =  \left[ \mathds{I} + r (\mathds{J}- r \mathds{I})^{-1} + r (\mathds{J}- r \mathds{I})^{-2} \left(\frac{ e^{\mathds{J} t_i } e^{-r t_i}-e^{\mathds{J} t_f } e^{-r t_f}}{t_f-t_i} \right) \right] \mathbf{c}_0.
\end{equation}
Thus, if the eigenvalues of $\mathds{J}$ have negative real part less $r$, then $\mathbf{C}_0 \approx \left[ \mathds{I} + r (\mathds {J}- r \mathds{I})^{-1}\right] \mathbf{c}_0$ after the transient time.

For linear and quadratic growth with $r>0$, if the eigenvalues of $\mathds{J}$ have a negative real part, after a transitory time, the deviations tend to zero, and therefore Therefore, for $ t\gg 0$,

\begin{equation} \label{eq:repconstante}
\mathbf{C}_0 \approx \mathbf{c}_0,
\end{equation}
and it is safe to take the fixed-point concentrations for sufficiently long times.

For sinusoidal growth, if the eigenvalues of $\mathds{J}$ have negative real parts far from zero, the periodic functions of $\boldsymbol{\delta C}(t)$ average zero over all intervals that are multiples of $2 \pi/\omega$. Thus, in this case, for low values of $|r|$ and a time interval consisting of whole periods, the  approximation as in \eqref{eq:repconstante} can also be used.

In this section we have used the assumption that concentration changes are negligible over an interval. This will depend on the type of growth, the rate of growth, the chemical reaction, and the proximity to the bifurcation. In the next section we will propose some criteria and exemplify their use for a specific example, showing through numerical simulations the scope of these approximations for the Brusselator.

\section{A study case: The Brusselator}\label{sec:brusselas}
To exemplify the predictions of our approximations studied previously, here we study the homogeneous state of the Brusselator in the equations. \eqref{eq:bruselas1}, which constitute an example of a reaction system where the fixed point is not the origin. This system has been studied in many references and, for the parameters used, we will follow the analysis of Ref. \cite{ledesma2020eckhaus}.

 The Brusselator has a single fixed point at $\mathbf{c}_0=(A,B/A)$ and performs a stability change via a Hopf bifurcation at $B=B_H=1+A^2$, where the state of the system changes to a limit cycle of linear frequency $\omega_0\approx A$ . The Turing conditions for a domain of fixed size require a low value for the ratio of diffusion coefficients, $\sigma=D_u/D_v$. For example, if $A=1$ and $\sigma=0.1$, the Turing bifurcation occurs at $B_T=1.732$, and the Turing region is between $B_T \leq B < B_H$. For our study, we will use these parameters and take different values of $B$ to emphasize the role of the bifurcation distance, expecting that the Turing conditions for the growing domain change slowly as a function of the parameter $r$.

\subsection{Some features of the homogeneous state}

\begin{figure*}[hbtp]
\centering
\includegraphics[width=0.8 \textwidth]{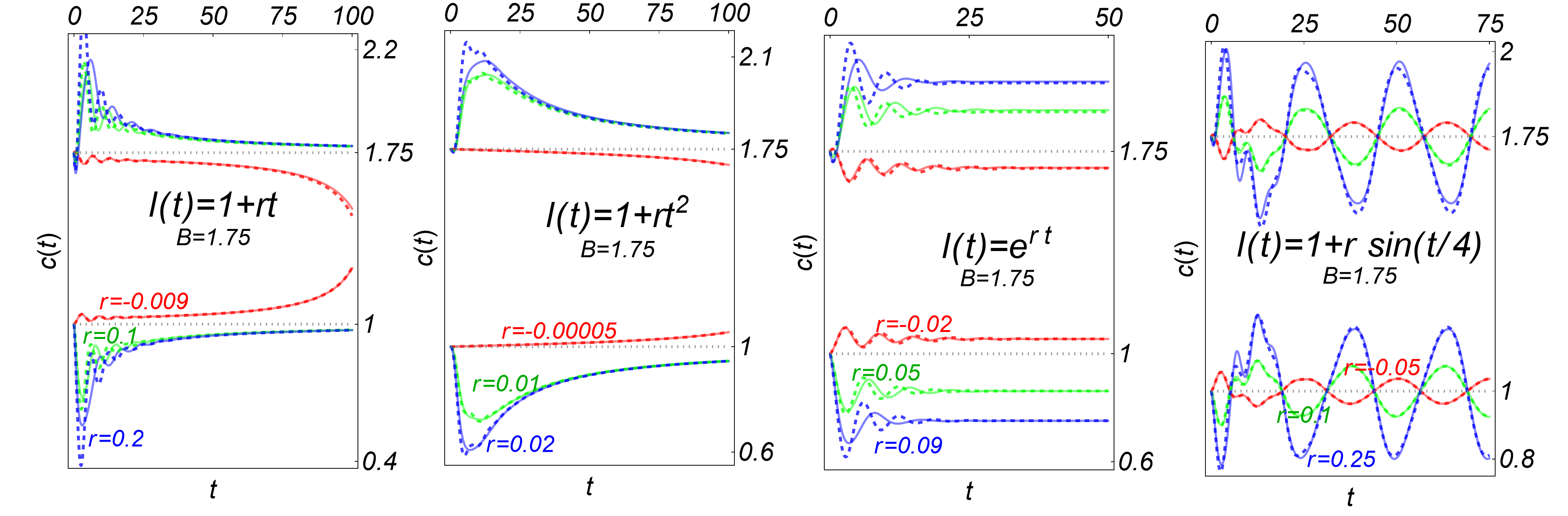}
\caption{Homogeneous state of the Brusselator with $B=1.75$. Each column represent different types of growing given in the inset. Solid lines: direct numeric solution of \eqref{eq:homogeneous}. Dashed lines: linear approximation in \eqref{eq:timeaproxima}. The different values of the parameter $r$ are represented by colors, blue and green for growing and red for shrinking. The fixed point concentrations are dotted lines in gray color.
 \label{fig:b1p75}}
\end{figure*}

In Fig. \ref{fig:b1p75}, we plot the numerical solutions of the homogeneous state solution of eq. \eqref{eq:homogeneous} for the reactions in \eqref{eq:bruselas1} with $B=1.75$. This value satisfies $B \ll B_H$, and therefore, in a fixed domain ($r=0$) it is expected that the solutions are stable without oscillations and that the concentrations tend to $(A,B/A)$. We show the behavior of the solution for both concentrations $(c_u,c_v)$ in solid lines with colors distinguishing three different values of $r$, red for shrinkage and green and blue for growth. Fixed-point concentrations are the gray horizontal dotted lines and their value is used for the initial condition. In each of the four columns of Fig. \ref{fig:b1p75} we plot the homogeneous state for the different types of growth studied in this work: linear, quadratic, exponential and periodic, respectively. The dashed curves show the same homogeneous state, but as predicted by our linear approximation in \eqref{eq:timeaproxima}, using the same colors for each value of $r$.

The results in Fig. \ref{fig:b1p75} show that for linear and quadratic growth (first and second columns of Fig. \ref{fig:b1p75}), the solutions present an initial outburst, and then tend asymptotically and slowly to the fixed-point concentrations $\mathbf{c}_0$, slower for quadratic than for linear. The amplitude of this burst seems proportional to the value of $r$. For linear and quadratic shrinkages, the initial solutions remain close to $\mathbf{c}_0$ at first, and, as the domain sizes approach zero, the concentrations begin to move away from the fixed point value by growing indefinitely.

For exponential growth/shrinkage (third column of Fig. \ref{fig:b1p75}), after a brief initial burst, solutions tend asymptotically to a value other than $\mathbf{c}_0$. The distance between the asymptotic and fixed-point concentrations grows with the absolute value of $r$. This tendency to the asymptotic value seems to occur faster than with the other types of growth studied.

In the case of sinusoidal growth (fourth column of Fig. \ref{fig:b1p75}), the size change produces concentration oscillations with the same frequency with which the domain changes, $\omega$. The amplitude of the oscillations is proportional to the absolute value of $r$. After an initial transient state, where several frequencies can be identified, the oscillations occur periodically with a single period $2\pi/\omega$.

\begin{figure*}[hbtp]
\centering
\includegraphics[width=0.8 \textwidth]{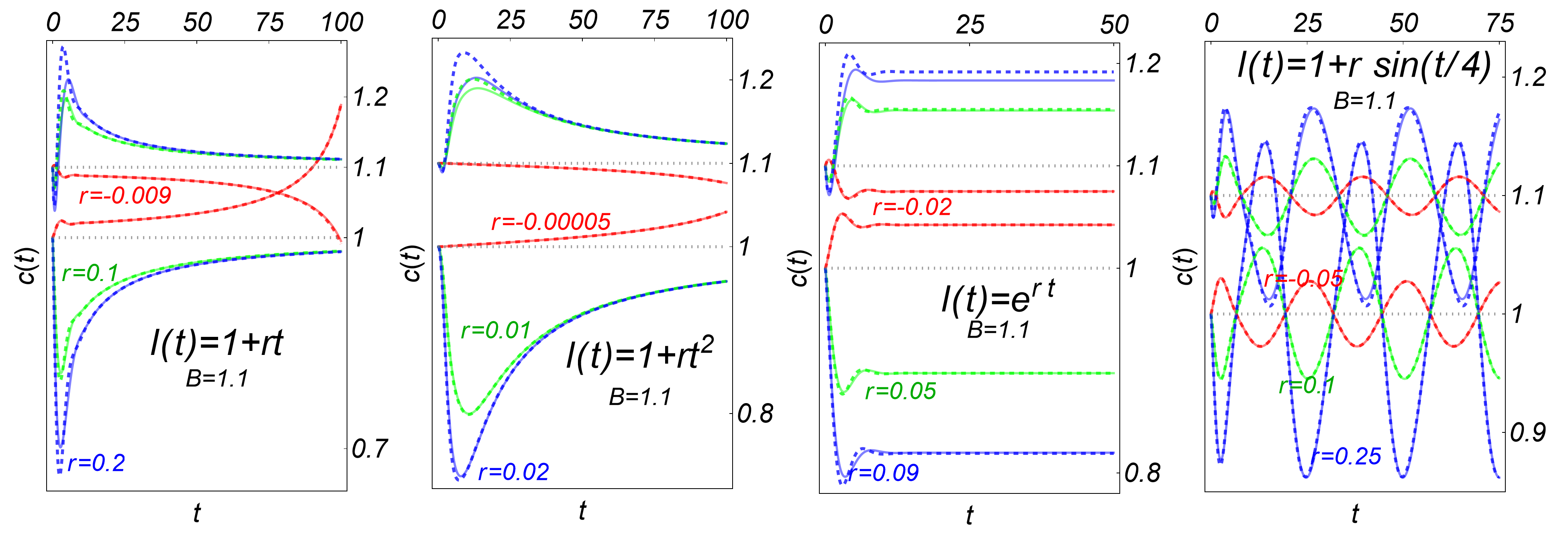}
\caption{Homogeneous state of the Brusselator with $B=1.1$. Same instructions as Fig \ref{fig:b1p75}. \label{fig:b1p1} }
\end{figure*}

\begin{figure*}[hbtp]
\centering
\includegraphics[width=0.8 \textwidth]{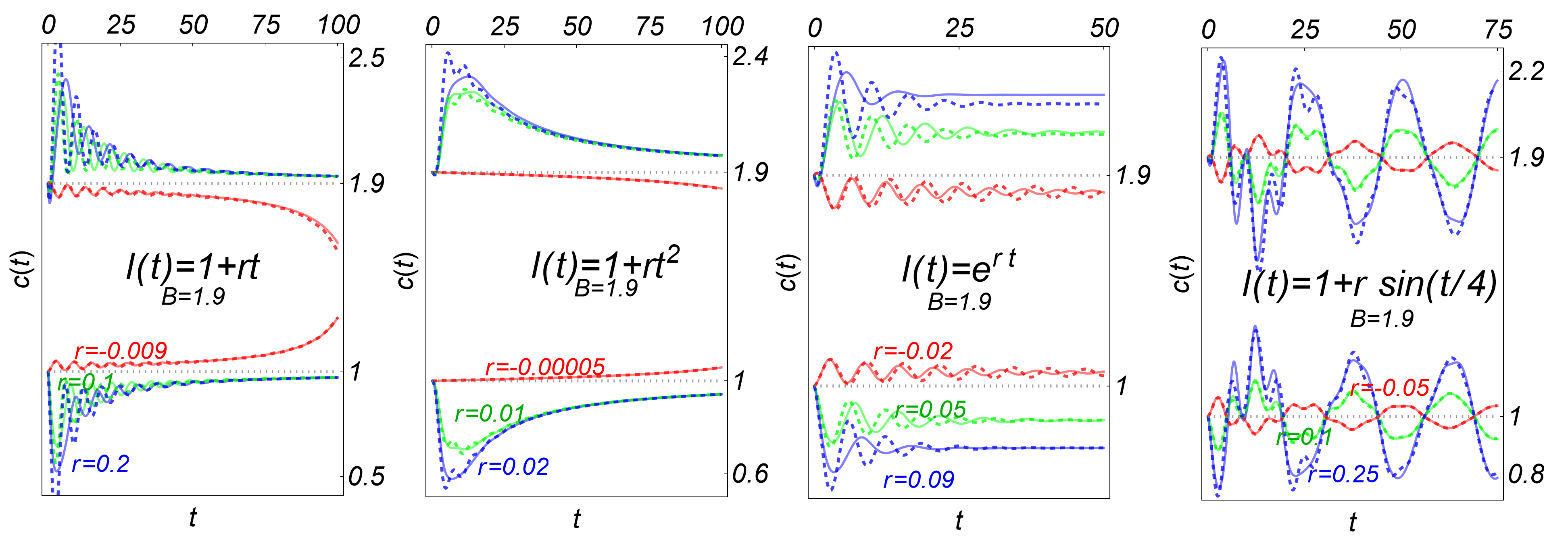}
\caption{Homogeneous state of the Brusselator with $B=1.9$. Same instructions as Fig \ref{fig:b1p75}. \label{fig:b1p9} }
\end{figure*}

These statements summarize the main variations of the homogeneous state concerning the parameter $r$. Now, in the top panel of Figs. \ref{fig:b1p1} ($B=1.1$) and \ref{fig:b1p9} ($B=1.9$), we repeat the results of \ref{fig:b1p75} ($B=1.75$) for other values of $B$ to test the effect of the distance to the Hopf bifurcation that occurs at $B_H=2$.

As can be seen from comparing  Figs. \ref{fig:b1p1} and \ref{fig:b1p9},  for the case of linear and quadratic growth, the amplitude of the initial burst increases with the parameter $B$, \emph{i.e.}, when the distance to the bifurcation decreases; in addition,  the number of initial oscillations and their duration is greater; however, the asymptotic trend to steady concentrations seems to occur faster for values of $B$ closer to $B_H$. For the exponential case, the steady state is reached earlier for values less than $B$, and the initial deviation from the fixed-point concentration increases with $B$. For sinusoidal growth, the amplitude of the oscillations around the fixed-point concentrations grow as $B$ increases.

\subsection{Validity of the time-dependent approximation in \eqref{eq:timeaproxima}}
As the Eq. \eqref{eq:timeaproxima} is a linear approximation to the nonlinear problem in \eqref{eq:homogeneous}, it is expected to be appropriate when the system is far from linearly unstable states and for small values of the dilution term. The first of these conditions requires values of $B$ smaller than the value of the Hopf bifurcation at $B_H=2$, and the second condition implies that the value of the dilution term proportional to $r$ is small (in absolute value). This is corroborated in all the comparisons made between the direct numerical solution for the homogeneous state (solid lines) and its analytical prediction (dashed lines) in Figs. \ref{fig:b1p75}-\ref{fig:b1p9}.  The scope of these two limitations of the model can be quantified as follows.

\begin{figure*}[hbtp]
\centering
\includegraphics[width=0.40 \textwidth]{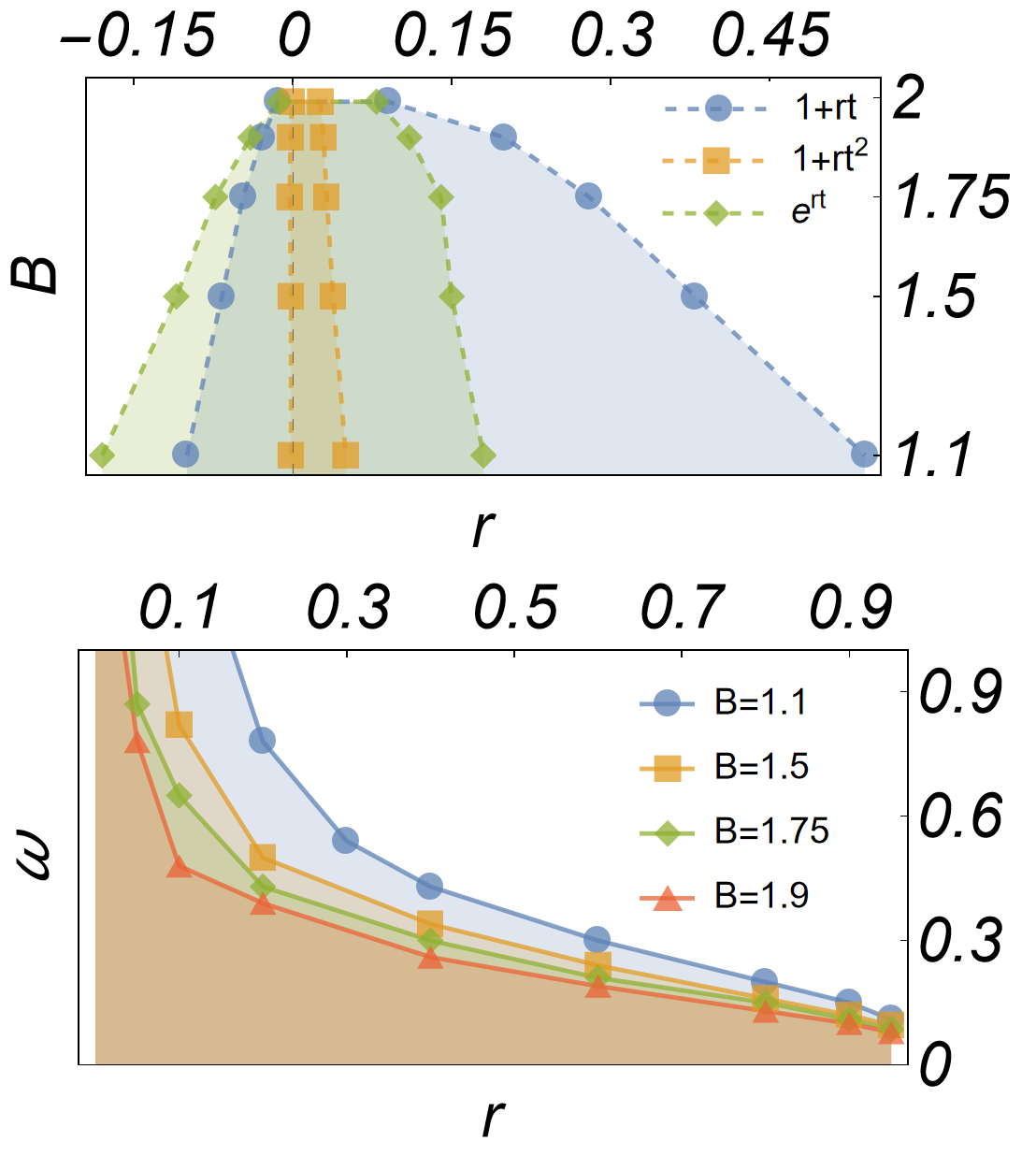}
\caption{Regions of validity of the approximation \eqref{eq:timeaproxima} to the solution of \eqref{eq:homogeneous} according to the criterion in \eqref{eq:error1}. At the top are considered the types of growth in the inset and at the bottom a periodic change in the size of the domain. \label{fig:validitylinear}}
\end{figure*}

In Fig. \ref{fig:validitylinear}, we plot the validity curves of \eqref{eq:timeaproxima} for the Brusselator model and the different types of growth. For each type of growth, we compare the numerical solution of \eqref{eq:homogeneous} with the solution in \eqref{eq:timeaproxima} by varying the growth parameter $r$. For linear, quadratic, and exponential growth, we increase the value of $r$ from zero until the solutions differ by more than 10\% under an error criterion described below; this error is time-averaged from $0$, up to a maximum time $t_{max}$ defined by the condition $l(t_{max})/l(0)=10$, \emph{i.e.}, the time required for the domain to grow to ten times its original length. For linear, quadratic, and exponential shrinkage, we do the opposite: we decrease the value of $r$ from zero until the solutions differ by more than 10\%, from time zero to a maximum time $l(t_{max})/ l(0) =1/10$, \emph{i.e.}, the time required for the domain to decrease to ten times its original length. The results of this type of comparison are given in Fig. \ref{fig:validitylinear}.a, where for each value of $B$, the two symbols express the maximum and minimum values of $r$ where \eqref{eq:timeaproxima} is valid for each type of growth. As can be seen in Fig. \ref{fig:validitylinear}.a, Eq. \eqref{eq:timeaproxima} is valid for low values of $|r|$ and its range of applicability decreases with distance from Hopf bifurcation. In addition, the asymmetry in the profile shows a greater width of applicability for growth than for shrinkage.

For sinusoidal growth, there are two parameters, $r$ and $\omega$. Therefore we modify both parameters starting both at $0$ until the numerical solution of \eqref{eq:homogeneous} and \eqref{eq:timeaproxima} vary $10\%$. The criterion for the maximum time used by us is that the system oscillates six times, or $t_{max}=6\cdot (2\pi/\omega)$. The results of this validation are given in Fig. \ref{fig:validitylinear}.b, where it is observed that the approximation has a wider range of validity for values of $B$ far from the Hopf bifurcation. Furthermore, it is observed that there is some kind of inverse relationship between the values of $r$ and $\omega$, which means that for high values of $r$ (great constriction), the periodicity of the changes in domain size has to be slow; and for low values of $r$ (low domain constraint), domain change frequencies can be higher.

To carry out this quantification, a comparison scheme must be defined that allows the difference between the two solutions to be measured. This depends on the time interval of interest in each system in question, which crucially depends on whether or not one is interested in transient times and on how large the domain ultimately becomes. For our comparisons, we choose the initial time as 0 and the final time as $t_{max}$ with the criteria explained above, \emph{i.e}, considering the process from the beginning. The error used in the comparisons in Fig. \ref{fig:validitylinear} is:

\begin{equation}\label{eq:error1}
error=\frac{1}{t_{max}} \int\limits_0^{t_{max}} \frac{|\mathbf{c}(t')-\mathbf{c}_s(t')|}{|c_0|}dt' < 10\%
\end{equation}
The numerator consists of the direct difference between the concentrations of \eqref{eq:homogeneous} and \eqref{eq:timeaproxima}, and the denominator provides the factor that allows us to compare with the values in the absence of growth. This criterion is arbitrary and can be adapted to the needs of each system. However, the results of Fig. \ref{fig:validitylinear} allow us to have a notion of when the expected homogeneous state can be predicted by \eqref{eq:timeaproxima}.

It is worth noting that the values of $r$ where our approximation is applicable express a wide range of growth/decrease situations. For example, for exponential growth, the value of $r$ is directly $g(t)=l'(t)/l(t)$; in finite difference terms this means $\frac{\delta l(t)}{l(t) \delta t}= r$, and therefore $r\times 100$ represents the percentage increase/decrease of the domain per unit of time. For example, for exponential growth with $B=1.75$, the scheme is valid for $r\sim 0.1$; this means that \eqref{eq:timeaproxima} approximates a system that increases up to ten times in each unit of time. To determine if this is fast or slow growth, for each situation one would have to compare the value of $r^{ -1}$ (which provides a characteristic time for growth) with the characteristic times for diffusion and reaction. The first is given by $D_i/l(t)^2$, where $D_i$ is the diffusion coefficients of the morphogens, and for the reaction, it is given by the inverse of the eigenvalues of the linear problem, $\mathcal{R}e[\Lambda_i]^ { -1}$. In general, for Turing patterns, the times required to grow in the formation of the biological pattern are much longer than those of the reaction-diffusion process and therefore, very low values of $r$ are required.

\subsection{Validity of the time-independent approximation in \eqref{eq:representativo}}

As we have shown before, the concentrations for exponential growth/shrinkage cases, for linear and quadratic growth for sufficiently long times, and sinusoidal domain variation with moderate amplitude changes, remain approximately constant throughout a characteristic value $\mathbf{C }_0$ found in \eqref{eq:representativo}. Therefore, in these cases, one can expect the Turing conditions, which in general would depend on $\frac{\partial \mathbf{f}}{\partial \mathbf{c}}(\mathbf{c}_s ) $, could be simplified to the approximation $\mathbf{c}_s(t) \approx \mathbf{C}_0$. If the nonlinear terms are small, it is expected that if both concentrations are similar, the variations in the above Jacobian matrix will also be small, since $\frac{\partial \mathbf{f}}{\partial \mathbf{c }}(\mathbf{c}_s ) \approx \frac{\partial \mathbf{f}}{\partial \mathbf{c}}(\mathbf{C}_0 )+\mathcal{O}[c_s(t )-\mathbf{C}_0] $. Therefore, in this section we will develop a criterion to know when this approximation is valid.

\begin{figure}[hbtp]
\centering
\includegraphics[width=0.40  \textwidth]{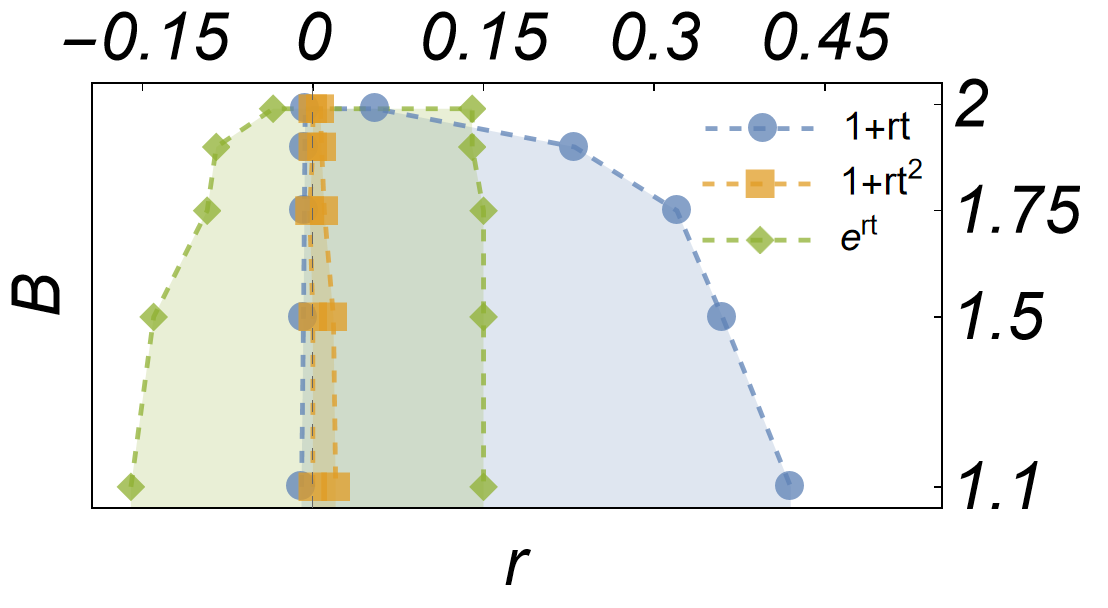}
\caption{Regions of validity of the approximation \eqref{eq:representativo} to the solution of \eqref{eq:homogeneous} according to the criterion in \eqref{eq:error2} for the growth functions in the inset. \label{fig:constante}}
\end{figure}

In Fig. \ref{fig:constante}, we test the validity of \eqref{eq:representativo} to approximate the direct numerical solution of the homogeneous state in \eqref{eq:homogeneous} by measuring its difference with the following error criterion

\begin{equation}\label{eq:error2}
error=\frac{1}{2\,t_{max}/3} \int\limits_{t_{max}/3}^{t_{max}} \frac{|\mathbf{c}(t')-\mathbf{C}_0|}{|c_0|}dt'< 5\%.
\end{equation}
This means we are considering only the last two-thirds of the process and the transient states up to $t_i=t_{max}/3$. As before, the choice of maximum time $t_{max}$ responds to a growth/shrinkage of ten times the original size of the domain.

The results of these comparisons in Fig. \ref{fig:constante} for the linear, quadratic, and exponential cases show that the approximation of \eqref{eq:repconstante} applies for low values of $|r|$ and over a wider range of  $B$ values far from the bifurcation. In contrast to the comparison made with the time-dependent solution \eqref{eq:timeaproxima} in Fig. \ref{fig:validitylinear}, the constant approximation \eqref{eq:repconstante} here is valid over a shorter range, with a more drastic reduction for a shrinking domain

For the case of oscillatory variation of the domain, we notice that the homogeneous state consists of oscillations around $\mathbf{c}_0$ with amplitude proportional to $r$ and therefore, if $n$ is the number of oscillations in time $t_{max}$, it is enough to have $n r/\omega$ small enough to satisfy this criterion. For this type of growth, we do not run simulations to test the criteria in \eqref{eq:error2} since it crucially depends on $n$.

\begin{figure}[hbtp]
\centering
\includegraphics[width=0.50  \textwidth]{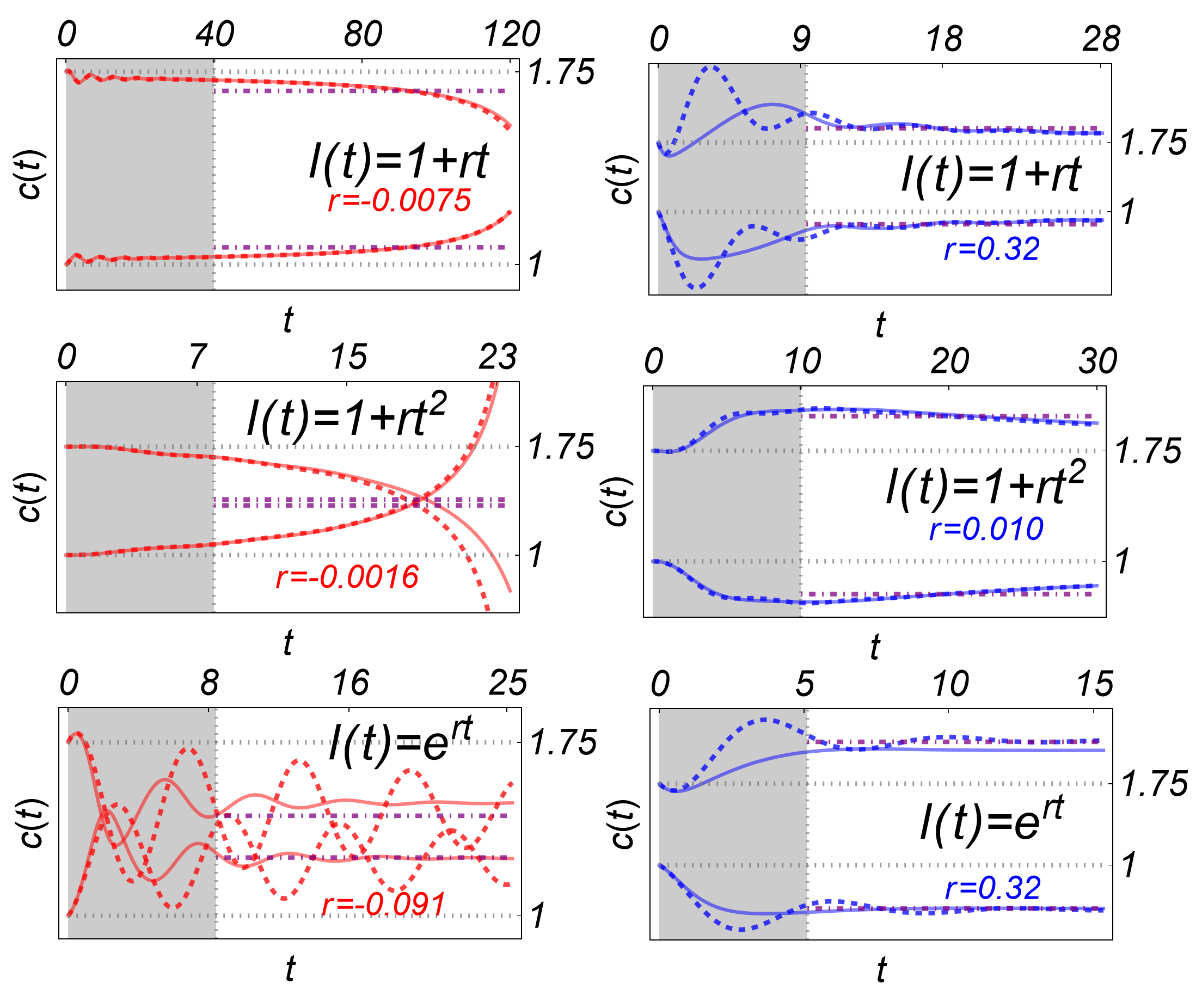}
\caption{ Comparisons of $\mathbf{c}_s$ in \eqref{eq:homogeneous} (solid lines), $\mathbf{c}_s$ in \eqref{eq:timeaproxima} (dashed lines), $\mathbf{C}_0$ in \eqref{eq:representativo} (magenta lines) and fixed-point concentrations $\mathbf{c}_0$ (gray dots).  The region in gray is the one excluded for the calculation of the last solution. \label{fig:constante2}}
\end{figure}

To better visualize the tolerance criteria defined in \eqref{eq:error1} and \eqref{eq:error2}, some comparisons of the three solutions of \eqref{eq:homogeneous} (solid lines), $\mathbf{c}_s$ in \eqref{eq:timeaproxima} (dashed lines) and $\mathbf{C}_0$ \eqref{eq:representativo} (magenta lines) as well as the fixed-point concentrations $\mathbf{c}_0$ (gray dots) are all plotted in Figure \ref{fig:constante2}. The parameters used for the simulations correspond to those where the maximum error is made according to both criteria. As can be seen, the similarity between the solutions in the interval defined by each criterion is acceptable.

It is worth noting that neither of the criteria set out in \eqref{eq:error1} and \eqref{eq:error2} can be general and are merely illustrative. The type of criteria to use in each case will depend on the time interval of interest for the formation of Turing patterns as well as the sensitivity of the system to changes in chemical concentrations due to dilution, as will be briefly discussed in the next section.

\section{Stability of the perturbations}\label{sec:perturbations}
We have shown that in an increasing domain, the homogeneous concentration $\mathbf{c}_s(t)$ changes with respect to the concentration it would have in a fixed domain, given by $\mathbf{c}_0$, constant. This means the stability properties of perturbations in \eqref{eq:perturbation} can change due to the dilution effect.

To quantify this, let's consider the perturbations in the absence of diffusion,  $\boldsymbol{\zeta}_0$, From Eq. \eqref{eq:perturbation}, they obey

 \begin{equation}\label{eq:perturbation0}
 \frac{\partial \boldsymbol{\zeta}_0}{ \partial t}   +\frac{\dot{l}(t)}{l(t)}\boldsymbol{\zeta}_0 = \frac{\partial \mathbf{f}}{\partial \mathbf{c}}(\mathbf{c}_s) \boldsymbol{\zeta}_0.
\end{equation} 
This equation depends explicitly on time through $\mathbf{c}_s(t)$. To make analytical progress, we can consider a time interval where $\mathbf{c}_s(t) \approx \mathbf{C_0}$, as we validated for the types of growth and parameters considered in Fig. \ref{fig:constante}. This approach allows solving \eqref{eq:perturbation0} directly in a similar way to the \ref{sec:homotime} section.

We multiply \eqref{eq:perturbation0} by $l(t)$ and make the approximation $ \frac{\partial \mathbf{f}}{\partial \mathbf{c}}(\mathbf{c }_s) \approx \frac{\partial \mathbf{f}}{\partial \mathbf{c}}(\mathbf{C}_0)$. To simplify notation, we define this matrix as $\hat{\mathds{J}}$, a constant Jacobian matrix, but evaluated at $\mathbf{C}_0$. This leads to

 \begin{equation}\label{eq:nueva2}
 \frac{\partial [l(t) \boldsymbol{\zeta}_0]}{ \partial t}  = \hat{\mathds{J}}[ l(t) \boldsymbol{\zeta}_0].
\end{equation} 
Let $\hat{\mathds{P}}^{ -1}$ be the modal matrix of $\hat{\mathds{J}}$ and let $\hat{\boldsymbol{\Lambda}}$ be its diagonal eigenvalue matrix ; both satisfy the property $\hat{\boldsymbol{\Lambda}}=\hat{\mathds{P}}^{-1}\hat{\mathds{J}} \hat{\mathds{P}}$. Making $l(t) \boldsymbol{\zeta}_0=\hat{\mathds{P}}^{-1} \mathbf{y}$, it can be proved on the same lines of Section \ref{sec:homotime} that $\mathbf{y}=e^{\hat{\boldsymbol{\Lambda}}t} \mathbf{y}_0$, and therefore, that the disturbances in the absence of diffusion satisfy:

\begin{equation}\label{eq:perturbation1}
\boldsymbol{\zeta}_0(t)=\frac{l(0)}{l(t)}  \hat{\mathds{P}}^{-1} e^{\hat{\boldsymbol{\Lambda}}t}  \hat{\mathds{P}} \boldsymbol{\zeta}_0(0).
\end{equation}
In this equation, the eigenvalues are those of the Jacobian at $\mathbf{C}_0$ and not at $\mathbf{c_0}$, as in a fixed domain, and expresses approximately the behaviour of perturbations when the homogeneous state does not change significantly in a time interval.

\subsection{Stability in absence of diffusion on growing domains}

Going back to the Turing conditions, we can make the hypothesis that the perturbations must be stable in the absence of diffusion, and therefore, for a growing domain, it would mean that disturbances must be damped and tend to zero in the absence of diffusion. In a fixed domain, this condition requires that the real part of all eigenvalues contained in $\boldsymbol{\Lambda}$ be negative. However, in a growing domain, this condition applied on \eqref{eq:perturbation1} leads to the criterion:

\begin{equation}\label{eq:stableperturbations}
\lambda(t)\equiv \mathcal{R}e[\hat{\boldsymbol{\Lambda_i}}]- \frac{1}{t}\log \frac{l(t)}{l(0)}<0  
\end{equation}
for all the eigenvalues numbered by  $i$. This means that homogeneous perturbations will damp if the local production of the modified reaction (quantified by $e^{\hat{\boldsymbol{\Lambda_i}}}$) grows more slowly than the domain growth (quantified by $l(t)$), which is reasonable. This effect reflects a part of the dilution effect. The second effect is implicit in the modification of the eigenvalues $\boldsymbol{\Lambda_i} \to \hat{\boldsymbol{\Lambda}}_i$ due to the change of the homogeneous state $\mathbf{c}_0 \to \mathbf{C}_0$, reflecting that the dilution also changed the local response of the reaction at each position and therefore the stability of the local reaction itself. Finally note that the criterion for fixed domains is recovered when $l(t)=l(0)$ since in this case the dilution effect is zero and the concentration change is null ($\mathbf{C}_0=\mathbf{ c}_0$) making $\boldsymbol{\Lambda_i} = \hat{\boldsymbol{\Lambda}}_i$.

\vspace{0.5cm}

The approximated criteria for the stability of perturbations in \eqref{eq:stableperturbations} presents some differences respect the one in fixed domain:

\begin{itemize}
\item For a growing domain ($l(t)/l(0)>1$), dilution causes that states with stable parameters in the fixed domain to become more stable, \emph{i.e.}, decaying with faster damping to its steady state. In physical terms, the consumption in a reaction is more rapidly controlled towards a steady state due to the increase in volume. Also, in contrast to a fixed domain, where a positive eigenvalue necessarily reflects linear instability, for a growing domain, the dilution effect can make the solution stable. Thus, it may be the case that chemical production can be overwhelmed by dilution and, although the number of molecules produced continues to increase, the concentration decreases if the volume increases more rapidly.

\item For a shrinking domain where $l(t)/l(0)<1$, the consequences are the opposite, and in general the dilution can make the states less stable. This can mean first, that the chemical damping can become slower due the domain shrinking, and second, even that a stable perturbation in a fixed domain where $\mathcal{R}e[\Lambda_i]<0$ can growth with time if the domain is shrinking faster than the chemical damping, since the diminishing volume produces an augment in concentration.

\item However, it must be emphasized that dilution also changes the steady-state concentrations. As we have observed in our previous simulations, this change is more pronounced for large values of the parameter that measures the size change ($r$ in our case) and for chemical parameters close to the bifurcation ($B$ in our case). In physical terms, this means that the growth/shrinkage effect is also capable of changing the local properties of the reaction, making it possible to change the stability of perturbations from the homogeneous state.

\end{itemize}

\vspace{0.5cm}

\begin{figure*}[hbtp]
\centering
\includegraphics[width=0.75 \textwidth]{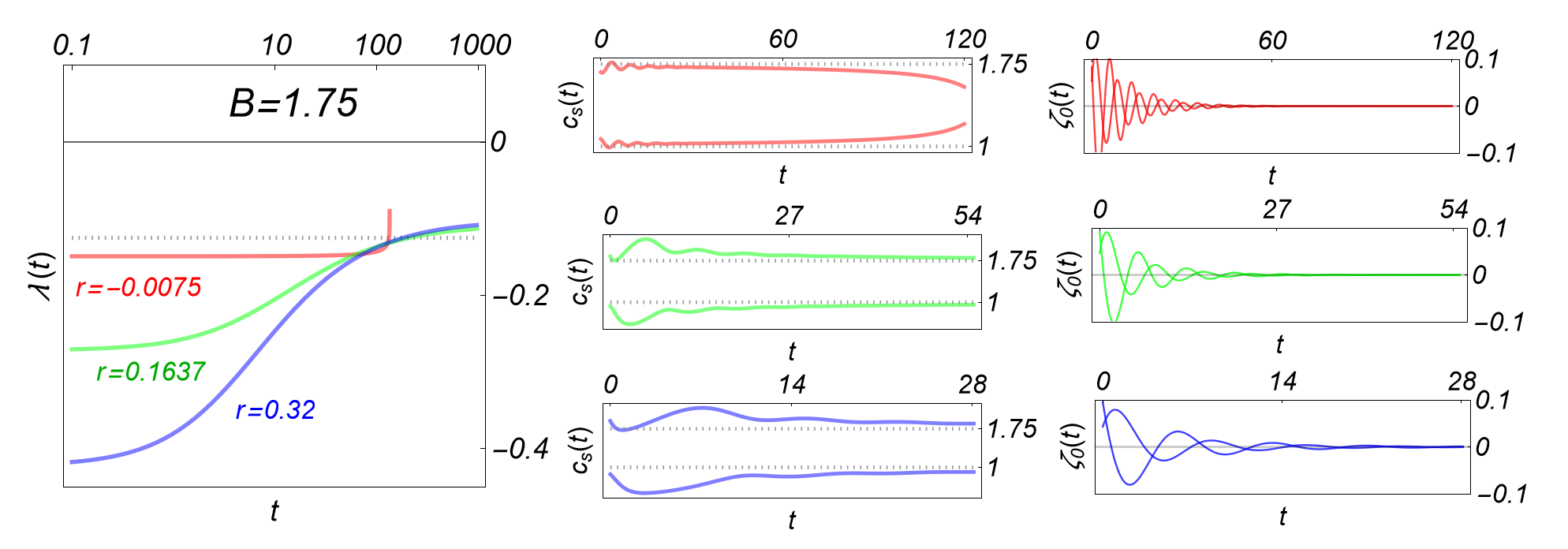}
\caption{Brusselator perturbations for $B=1.75$ and linear growth $l(t)=1+r t$. Left: Rate $\lambda(t)$ in \eqref{eq:stableperturbations} for each value of $r$ (red, blue, and green solid lines) and the maximum eigenvalue of the reactive system in a fixed domain (grey dots). Center and Right: numerical solutions for the homogeneous state, \eqref{eq:homogeneous} and perturbations \eqref{eq:perturbation1}, respectively for each of the three values of $r$ on the left with the same colors. \label{fig:eigenvalues}}
\end{figure*}

\begin{figure*}[hbtp]
\centering
\includegraphics[width=0.75 \textwidth]{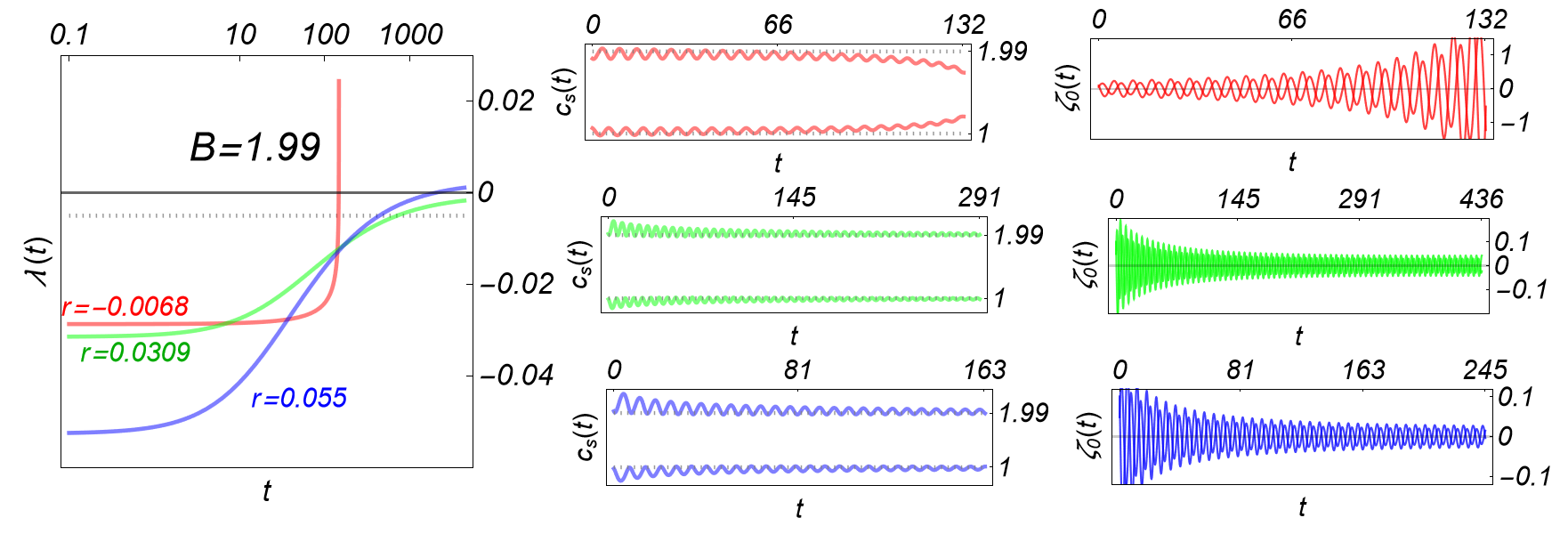}
\caption{ Brusselator perturbations for $B=1.99$ and linear growth $l(t)=1+r t$. Same instructions as Fig. \ref{fig:eigenvalues}. \label{fig:eigenvalues2}}
\end{figure*}

To illustrate these situations, in Fig. \ref{fig:eigenvalues} and \ref{fig:eigenvalues2} we plot the value of $\lambda(t)$ in \eqref{eq:stableperturbations} for linear growth $l( t)=1 + r t$ as a function of time for different values of $r$ represented by different colors, red, green and blue. The values of $r$ and $B$ lie within the region where the approximation of $\mathbf{C}_0$ in \eqref{eq:representativo} is valid by the criterion in \eqref{eq:error2}. The values of $B$ in Figs. \ref{fig:eigenvalues} and \ref{fig:eigenvalues2} correspond to $1.75$ and $1.99$, respectively.

In the first case of Fig. \ref{fig:eigenvalues}, we represent the most common situation we observed in most of our simulations, where the rate $\lambda(t)$ remains negative, even in shrinking domains($r<0$). In these cases, the stability of the state $\mathbf{c}_0$ is inherited by $\mathbf{C}_0$, even considering the effect of dilution and proximity to the bifurcation. This same negativity of $\lambda(t)$ was also observed for the other types of growth in the whole range of $r$ values considered  .

In Fig. \ref{fig:eigenvalues2}, we show a situation where there can be a change in stability due to the combined effect of high values of $|r|$ and the proximity of the Hopf bifurcation. For this choice of parameters, it is observed that the rate $\lambda(t)$ can start to grow until it is positive making the perturbation $\boldsymbol{\zeta}_0$ become unstable. This occurs for both the growing (in blue) and shrinking (in red) domains. To demonstrate that this change in stability in these cases actually occurs due to dilution and not as a result of the different approximations made in \eqref{eq:stableperturbations}, in Fig. \ref{fig:eigenvalues2}, we plot the perturbation \eqref{eq:perturbation0} using $\mathbf{c}_s(t)$ obtained directly from \eqref{eq:homogeneous}, and not from the approximate representative value, demonstrating the growth of concentrations predicted by $\lambda(t) $ in this section is adequate.

\section{Differences with a system with fixed-point concentration at the origin: the BVAM example}
\label{sec:bvam}

The ec. \eqref{eq:timeaproxima} allows us to predict that a system with the origin as the only fixed point ($\mathbf{c}_0=\mathbf{0}$) cannot change that value of its steady homogeneous concentration. To exemplify this, we can use the BVAM in \eqref{eq:bvam1} which, with coefficients $a=-1$ and $h=3$ and $-3<b<-1$, exhibits a single fixed point in the origin. The Hopf bifurcation occurs at $b_H=-1$, where the system has periodic oscillations \cite{ledesma2019primary}.

\begin{figure}[hbtp]
\centering
\includegraphics[width=1 \textwidth]{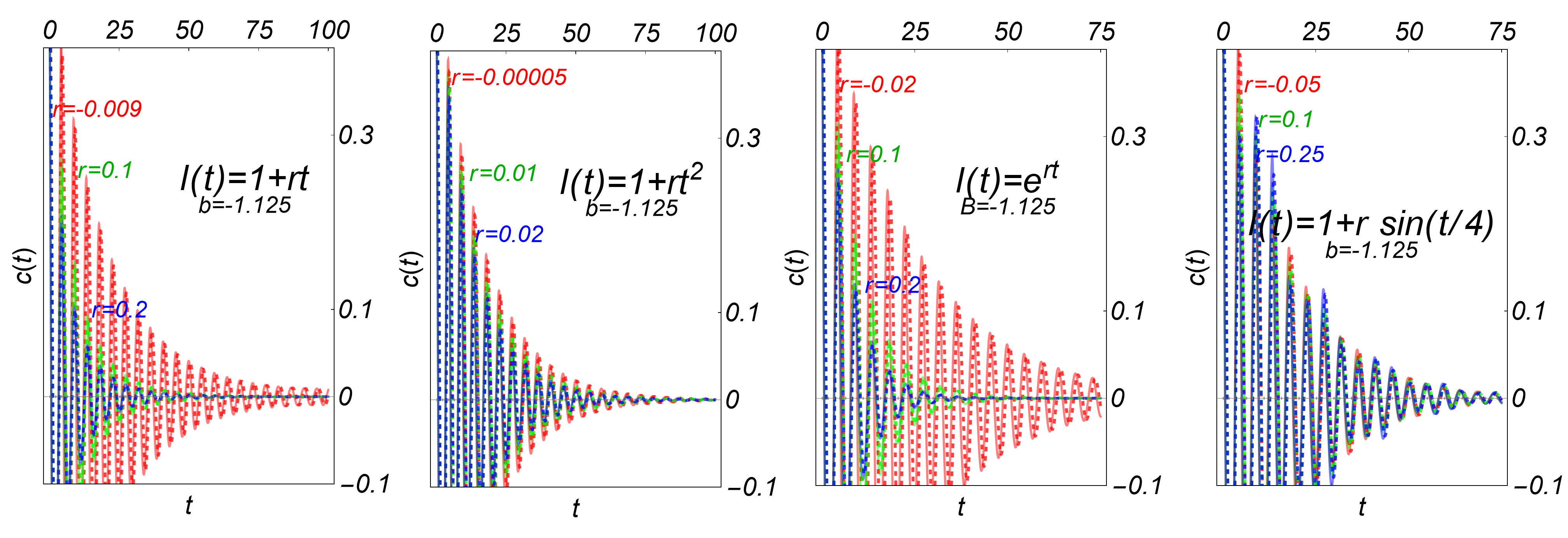}
\caption{Homogeneous state for the first concentration of BVAM with $b=-1.125$ using the direct solution of \eqref{eq:homogeneous} (solid lines) and the approximation in \eqref{eq:st1perturbation1} (dashed lines). The different types of growth are given in the inset of each column, and the values of $r$ are distinguished by color: growth (green and blue) and shrinkage (red). \label{fig:bvamsol}}
\end{figure}

In Fig. \ref{fig:bvamsol}, we plot the direct numerical solutions of the BVAM in \eqref{eq:bvam1} (solid lines) and the analytical approximation in \eqref{eq:st1perturbation1} (dashed), using an initial deviation given by $\delta \mathbf {c}(0)=(1,1)^T$, giving a good comparison. This means that our approximation is also valid for the BVAM with the parameters used.

Now we compare the type of solutions obtained from the BVAM and Brusselator systems. In the case of the Brusselator system in Fig. \ref{fig:b1p75} with fixed point at $(A,B/A)$ , the dilution can produce a constant change in the stationary concentration (exponential growth), oscillations around of a state of concentration other than zero (oscillatory) and an asymptotic and slow tendency to its fixed-point value (linear and quadratic cases). The BVAM system in Fig. \ref{fig:bvamsol} with a single fixed point at the origin has essentially the same asymptotic behaviour for all types of growth. This type of fundamental difference will allow us to identify which reaction system is more suitable for the object of study that is being modelled in each case.

\begin{figure}[hbtp]
\centering
\includegraphics[width=0.75 \textwidth]{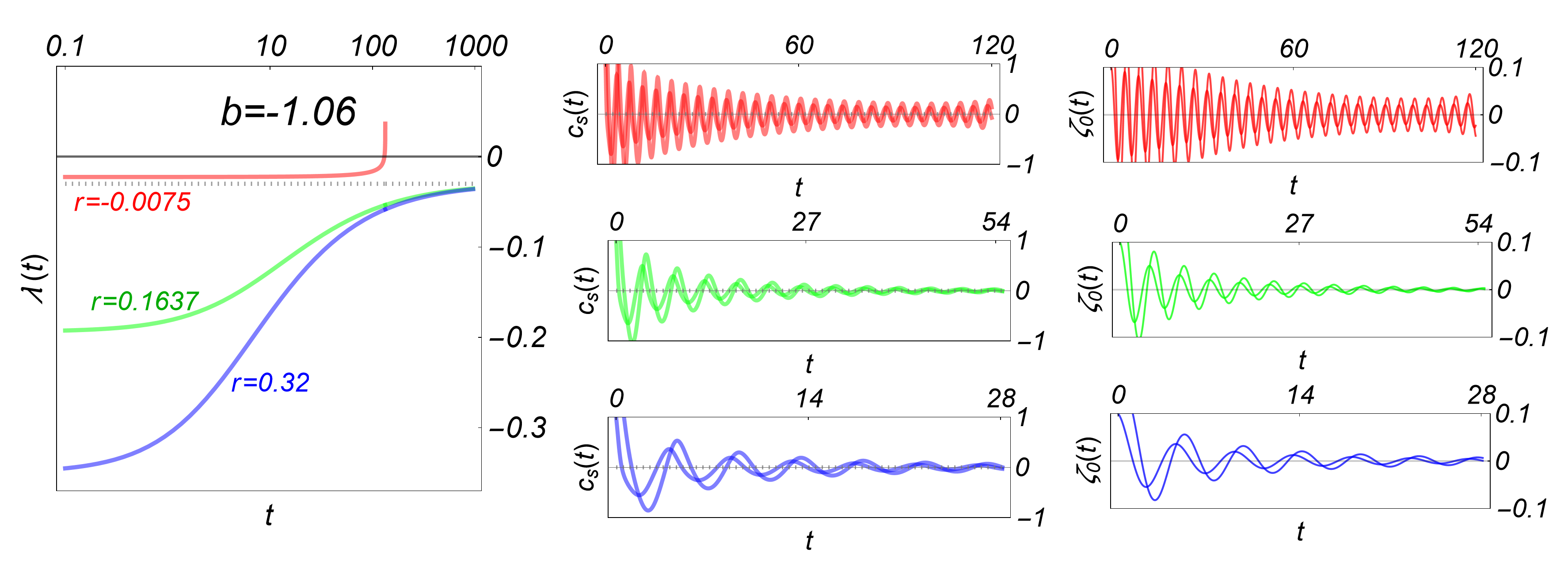}
\caption{BVAM perturbations for $b=-1.06$ and linear growth $l(t)=1+r t$. Same instructions as in Fig. \ref{fig:eigenvalues}. \label{fig:perturvbvam1}}
\end{figure}

Another aspect to highlight related to the position of the fixed point is related to the stability of the homogeneous state and perturbations. In the BVAM system, if the homogeneous state is asymptotically stable, according to \eqref{eq:stableperturbations}, the only way to destabilize the system corresponds to domain reduction. This is exemplified in Fig. \eqref{fig:perturvbvam1} where we repeat the analysis of Section \ref{sec:perturbations} but now for the BVAM, plotting the results of \eqref{eq:stableperturbations} the homogeneous state to the left and disturbances to the right. Since in this case $\mathbf{C}_0=\mathbf{0}$, then $\lambda(t)= \mathcal{R}e[\boldsymbol{\Lambda_i}]- \log [l (t)/l(0)]/t$, and therefore, the dilution effect can increase the value of $\lambda$ only if $l(t)<l(0)$. We exemplify this for linear growth in the BVAM system where the curves of $\lambda$ are necessarily below the eigenvalue (grey line) for growing (green and blue cases) and above to shrinking (red curve). This is in contrast to what occurs in Figs. \ref{fig:eigenvalues} and \ref{fig:eigenvalues2} for the Brusselator, where the curves of $\lambda(t)$ can be above and below the maximum eigenvalue. Therefore, the stability of the perturbation in fixed and growing domains also establishes differences that allow to choose different types of reaction systems.

\begin{figure}[hbtp]
\centering
\includegraphics[width=0.45 \textwidth]{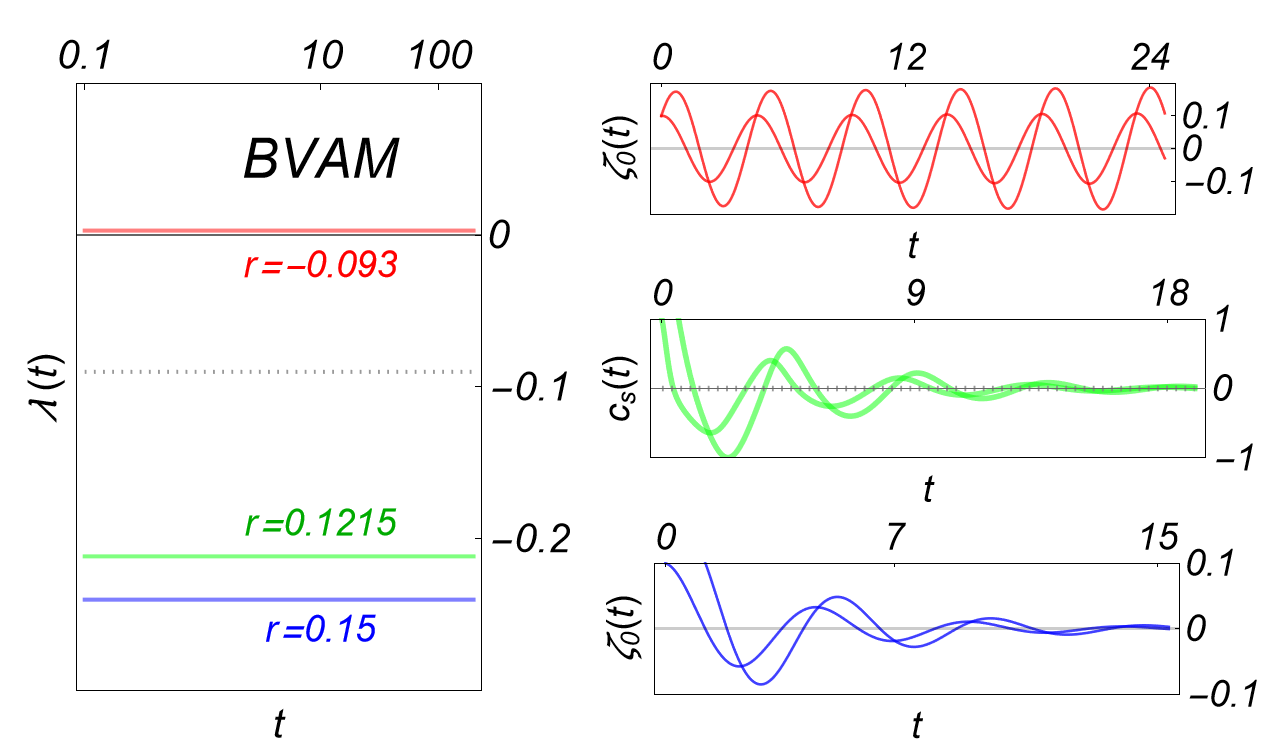}
\includegraphics[width=0.45  \textwidth]{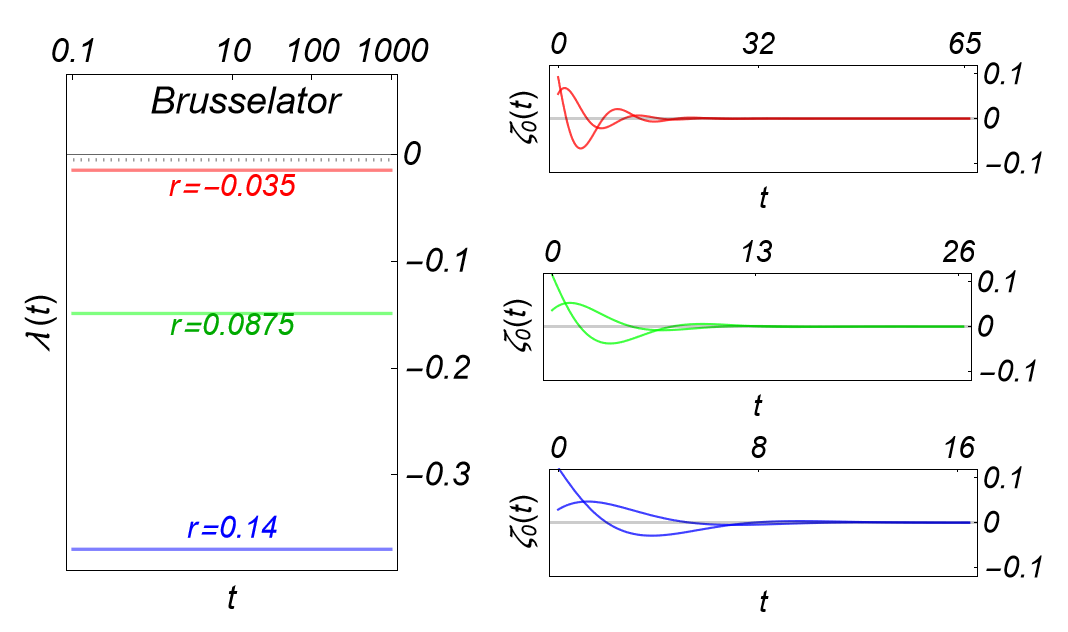} 
\caption{Behaviour of the perturbations for the BVAM with $b=-1.18$ and Brusselator with $B=1.99$ for exponential growth with $l(t)=e^{rt}$. We show the predicted rate $\lambda(t)$ of each perturbation at the left sides of each figure for three different $r$ parameter according to \eqref{eq:stableperturbations}  (in  red, green and blue colors) and the maximum eigenvalue of the fixed domain problem. At the right sides we plot the perturbations using the direct numerical solutions of \eqref{eq:perturbation0} with \eqref{eq:homogeneous}.   \label{fig:exponential}}
\end{figure}

One last aspect to comment on the differences between both systems refers to the stability of disturbances. In the case of BVAM, there is no change in steady concentration, and the change in size causes the net rate $\lambda(t)$ to be above or below the maximum eigenvalue associated with the reaction. This is directly illustrated for exponential growth in Fig. \ref{fig:exponential}.a, where the decreasing domain (in red colors) is capable of changing the stability of BVAM perturbations. In the case of the Brusselator, contrary to what might be expected for exponential growth, even for strong shrinkage, in our simulations, we observe that both the homogeneous state solution and the perturbations almost never lose stability. This is illustrated in Fig. \ref{fig:exponential}.b, where even for relatively large shrinkage (producing large differences between $\mathbf{c_0}$ and $\mathbf{C}_0$), both $ \mathbf{c}_s$ and $\boldsymbol{\zeta}_0$ tend asymptotically to a stationary value. In physical terms, this may suggest that the change in concentration of $\mathbf{c_0} \to \mathbf{C}_0$ caused by dilution is intended to preserve the stability of material production/consumption locally as the system grows or shrinks.

\section{Conclusions and discussion}\label{sec:discussion}

In this work, we have studied in detail the homogeneous state resulting from the reaction and diffusion processes in a one-dimensional increasing domain. We present a scheme that allows us to predict that the effects of the dilution term on the dynamics are the following: 1) It makes the homogeneous state dependent on time; 2) it is capable of making steady concentrations different from the expected fixed-point concentrations in a fixed domain, and 3) it can modify the stability properties of perturbations due to local volume change.

To predict the homogeneous state, we proposed a solution based on a linear approximation that, through numerical comparison  with the numerical solution, we demonstrated that it correctly incorporates the aforementioned characteristics of the homogeneous state. This approximation is validity in the linear regime, where the domain growth is slow and for systems where the reaction parameters are below the bifurcation. The applicability of our approximation was verified with an extensive comparison against numerical solutions for the homogeneous state for different types of growth: exponential, linear, quadratic, and oscillatory.

We show that the deviation from the steady state of the fixed-point concentrations occurs due to dilution, mainly in non-equilibrium systems where the fixed-point concentration of the reaction is non-zero. This stationary concentration change occurs especially for exponential growth. We show that after some time, a system with linear or quadratic growth tends to recover the concentrations that it would have in the absence of growth, while the oscillatory variation of the domain size produces periodic oscillations around the fixed-point concentrations.

It is worth mentioning that the importance and applicability of our predictions for the homogeneous state in the search for Turing patterns lies in the fact that, as we have commented, the stability of the perturbations is inherited in most cases from the stability of the stationary state. Thus, a stable steady state in a fixed domain will in most cases inherit its stability to perturbations. However, it should also be taken into account that the type of chemical system (position of the fixed point), as well as the shrinkage or the proximity to the bifurcation can modify the parameter space where the perturbations are stable, since the stability of the perturbations is modified by the dilution. To quantify this effect, we demonstrate a linear approximation for the growth of disturbances that, for the parameters and system studied,  proved to be effective in predicting perturbation stability, as corroborated with numerical simulations.

Regarding the Turing conditions for the appearance of spatial patterns, in this work, we study those related to stability in the absence of diffusion and establish that in an increasing domain, this stability depends on three factors: the change in the concentrations in the state stationary, the local volumes' change that affects the concentration, and the stability that the chemical reaction would have in the absence of growth. Competition between these factors provides richer conditions for pattern emergence than those found in a fixed domain where only the last of these factors matters.

The formal results given here for the homogeneous state and perturbation stability are general and can be applied to many-component chemical systems. Some of the theoretical proofs were explicitly confirmed in the form of matrix components for two-component systems. Our predictions were also tested against numerical simulations of the homogeneous state and perturbations for the Brusselator and BVMA reactions. Therefore, we believe that our results are robust and can be applied to predict temporal variations of the homogeneous state of general RDD systems and help in future works to establish the conditions for the formation of Turing patterns in growing domains when diffusion is included.

\bibliography{biblio1}

\end{document}